\documentclass[showpacs,preprintnumbers,eqsecnum]{revtex4}

\usepackage{amsmath}
\usepackage{makeidx}
\usepackage{graphicx}
\usepackage{bm}
\usepackage{hyperref}

\begin{document}

\preprint{}
\title{Applications of the electromagnetic field theory for calculations of the radiative thermal transfer in coaxial multi-layer systems}
\author{C. Stoica$^{1}$}
\affiliation{$^{1}$Department of Physics, University of Bucharest,
MG11, Bucharest-Magurele 76900,Romania}

\begin{abstract}
\textbf{Abstract} In the present article we intend to accomplish a
rigorous analysis which allows to elaborate useful symbolic and
numerical codes leading to an accurate evaluation of the thermal
radiation. We will consider the case of the electromagnetic field in
axially symmetric systems, a case of real interest in cryogenics.
\end{abstract}

\maketitle

\section{Electromagnetic field in conductive bounded cylindrical systems}

The first topic is the electromagnetic field inside a hollow
cylinder cavity with conducting walls and the spectrum of resonance
frequencies associated to the various modes of the field.

We will assume that the field has a monochromatic time dependence
$e^{i\omega t} $. The symmetry axis will be taken as the z-axis of
the coordinate system. With this choice, the field will split in two
orthogonal components as
    \begin{equation}\vec E = \vec e_z E_z  + \vec E_
    \bot\end{equation}

The walls of the cavity are situated at $\rho  = b$, $z = 0$ and $z
= l$.

The tangential component of the field should vanish on the
conducting walls, that is $E_z  = E_\phi   = 0$ on the cylinder
surface $\rho  = b$ and $E_\rho   = E_\phi   = 0$ at $z = 0{\rm{ and
}}z = l$ (fig.1). Inside the hollow cavity the field will obey the
wave equation
    \[\left( {\nabla ^2  - \frac{1}{{c^2 }}\frac{{\partial ^2 }}{{\partial t^2 }}} \right)\vec E =
    0\]

or
    \[\left( {\nabla ^2  + \frac{{\omega ^2 }}{{c^2 }}} \right)\vec E =
    0\]

for the case of harmonic field of frequency $\omega $. The axial
component of the space part of the electric field obeys the same
equation:
    \begin{equation}\left( {\nabla ^2  + \frac{{\omega ^2 }}{{c^2 }}} \right)E_z (\rho ,\phi ,z) = 0\end{equation}

and so does the transverse part $\vec E_ \bot  :\left( {E_\rho
,E_\phi  } \right)$. The equation can be written in cylindrical
coordinates as
     \begin{equation}\left( {\frac{{\partial ^2 }}{{\partial \rho ^2 }} + \frac{1}{\rho }\frac{\partial }{{\partial \rho }} + \frac{1}{{\rho ^2 }}\frac{{\partial ^2 }}{{\partial \phi ^2 }} + \frac{{\partial ^2 }}{{\partial z^2 }} + \frac{{\omega ^2 }}{{c^2 }}} \right)E_z  = 0\end{equation}

and assumes a separate solution of the type \[E_z (\rho ,\phi ,z) =
R(\rho )\Phi(\phi )Z(z)\]. Thus, the Helmholtz PDE above becomes
    \[\frac{1}{R}\left( {\frac{{d^2 R}}{{d\rho ^2 }} + \frac{1}{\rho }\frac{{dR}}{{d\rho }}} \right) + \frac{1}{{\rho ^2 \Phi}}\frac{{d^2 \Phi}}{{d\phi ^2 }} + \frac{1}{Z}\frac{{d^2 Z}}{{dz^2 }} + \frac{{\omega ^2 }}{{c^2 }} =
    0\]

or
     \begin{equation}\frac{1}{R}\left( {\frac{{d^2 R}}{{d\rho ^2 }} + \frac{1}{\rho }\frac{{dR}}{{d\rho }}} \right) + \frac{1}{{\rho ^2 \Phi}}\frac{{d^2 \Phi}}{{d\phi ^2 }} + \frac{{\omega ^2 }}{{c^2 }} =  - \frac{1}{Z}\frac{{d^2 Z}}{{dz^2 }} = k^2\end{equation}

    The reflections at the top $z = l$and bottom $z = 0$ surfaces impose a z-dependence appropriate for standing waves. Thus, in the eq. (1.6) the constant $k^2  \ge 0$
and the solution $Z(z)$ will be a combination of  $\sin kz{\rm{ and
}}\cos kz$ which will obey the boundary conditions. Further, in
order to have a uniform $\phi $
 dependence, $\Phi(\phi )$ should be of the form $\Phi(\phi ) = e^{ \pm im\phi } $ with integer $m = 0,1,2,3,...$

    Thus, the radial equation becomes
     \begin{equation}\frac{{d^2 R}}{{d\rho ^2 }} + \frac{1}{\rho }\frac{{dR}}{{d\rho }} + \left( {\gamma ^2  - \frac{{m^2 }}{{\rho ^2 }}} \right)R = 0\end{equation}

where  $\gamma ^2  = \frac{{\omega ^2 }}{{c^2 }} - k^2 $. The
equation (1.7) is the Bessel equation of order $m = {\rm{integer}}$
and the solution is a combination of Bessel and Neumann functions of
order m , $J_m (\gamma \rho ){\rm{ and }}N_m (\gamma \rho )$. If we
take into account that the solution should be finite for any $\rho
$, including $\rho  = 0$ and remember the behavior of Bessel and
Neumann functions in the vicinity of the origin
    \[J_m (x) \to \frac{1}{{\Gamma (m + 1)}}\left( {\frac{x}{2}}
    \right)^m\]

respectively
     \[N_m (x) \to  - \frac{{\Gamma (m)}}{\pi }\left( {\frac{2}{x}}
     \right)^m\]

for $m \ge 1$ and
     \[N_0 (x) \to \frac{2}{\pi }\left[ {\ln \left( {\frac{x}{2}} \right) + 0.5772...}
     \right]\]

than only $J_m (\gamma \rho ){\rm{ }}$ is acceptable, because the
second solution is singular at the origin.
    In the case of Transverse Magnetic (TM) modes $H_z  = 0$ everywhere in the cavity,
    while the transverse components of the electric and magnetic
    field are
     \begin{equation}\vec E_ \bot   = \frac{1}{\gamma }\nabla _ \bot  \frac{{\partial E_z }}{{\partial z}}\end{equation}

and, respectively
     \begin{equation}\vec B_ \bot   = \frac{{i\omega }}{{c^2 \gamma ^2 }}\vec e_z  \times \nabla _ \bot  E_z\end{equation}

where $\nabla _ \bot  :\left( {\frac{\partial }{{\partial
\rho}},\frac{1}{\rho }\frac{\partial }{{\partial \phi }}} \right)$
is the Hamilton's operator in polar coordinates.

\begin{figure}
\centering
\includegraphics[width=3in,keepaspectratio=true]{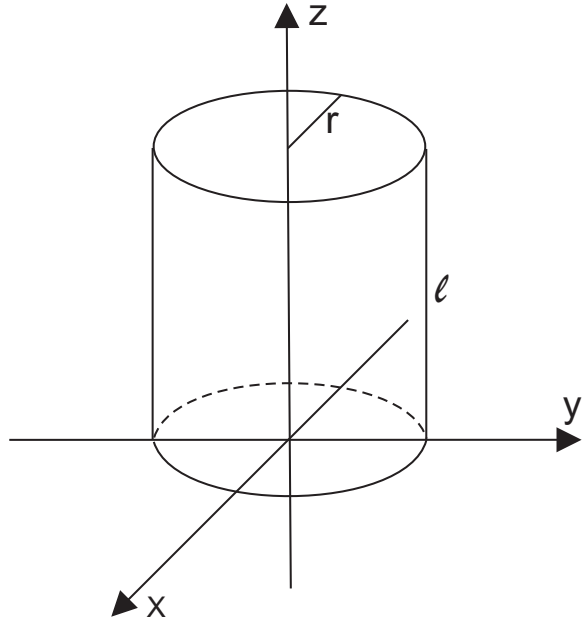}
\caption{The geometric parameters of the system}
   \label{fig.1}
\end{figure}

Than $E_z $ must assume the form
     \begin{equation}E_z (\rho ,\phi ,z) = E^0 J_m (\gamma \rho )e^{ \pm im\phi } \cos \left( {p\pi \frac{z}{l}} \right){\rm{ with }}p = 0,1,2,...\end{equation}

in order that $\vec E_ \bot$ to satisfy the boundary conditions at
$z = 0{\rm{ and }}z = l$. Also, the boundary condition  $E_z  = 0$
for $\rho  = b$ requires that the argument $\gamma b$ should be one
of the zeros of the Bessel function $J_m$. Thus, $\gamma $ will
assume only a set of discrete values
     \[\gamma _{mn}  = \frac{{x_{mn} }}{b}\]

where $x_{mn}$ is the n-th zero of $J_m $.

    The general solution of the field equations (1.10) inside the cavity that obeys the appropriate boundary conditions will be a superposition of the normal modes :

     \begin{equation}E_z (\rho ,\phi ,z) = \sum\limits_{m = 0}^\infty  {\sum\limits_{n = 1}^\infty  {\sum\limits_{p = 0}^\infty  {E_{mnp} J_m (x_{mn} \frac{\rho }{b})e^{ \pm im\phi } \cos \left( {p\pi \frac{z}{l}} \right)} } }\end{equation}

    The resonance frequencies of the cavity will depend on three parameters and will be given in the case of TM modes by
    \begin{equation}\omega _{mnp}  = c\sqrt {\left( {\frac{{x_{mn} }}{b}} \right)^2  + \left( {\frac{{p\pi }}{l}} \right)^2 }\end{equation}

with m=0,1,2,...  n=1,2,...  p=0,1,2,...

\section{Hollow cylindric ring with conducting walls}

The second topic is the electromagnetic field inside a hollow cavity
of the form of a cylindric ring with conducting walls, with inner
radius $a$, outer radius $b$and height $l$.

The PDE and the boundary conditions that the fields obey will be the
same as in the case of the cylinder cavity, except that a
supplementary boundary at $\rho  = a{\rm{ }}( > 0)$ appears. Since
the origin $\rho  = 0$ no longer belongs to the domain of values
that $\rho $ assumes, than the particular solution of the radial
equation will be now a linear combination of both Bessel and Neumann
functions:
    \begin{equation}R_m(\rho ) = A_m J_m (\gamma \rho ) + B_m N_m (\gamma \rho )\end{equation}

    In order that the field to satisfy homogenous boundary conditions on both $\rho  = a$and $\rho  = b$ boundaries, the radial funtion above should assume the form
    \begin{equation}R(\rho ) = A\left[ {J_m (\gamma _{mn} \rho ) - \frac{{J_m (\gamma _{mn} a)}}{{N_m (\gamma _{mn} a)}}N_m (\gamma _{mn} \rho )} \right]\end{equation}

where $\gamma _{mn} {\rm{   ,  }}n = 1,2,3,...$
 are the solutions of the equation
    \[J_m (\gamma _{mn} b)N_m (\gamma _{mn} a) - J_m (\gamma _{mn} a)N_m (\gamma _{mn} b) =
    0\]

They will be determined by appropriate numerical methods.

In the case of TM modes, the general solution for the axial
component $E_z $ is the linear combination:
    \begin{equation}E_z (\rho ,\phi ,z) = \sum\limits_{m = 0}^\infty  {\sum\limits_{n = 1}^\infty  {\sum\limits_{p = 0}^\infty  {E_{mnp} \left[ {J_m (\gamma _{mn} \rho ) - \frac{{J_m (\gamma _{mn} a)}}{{N_m (\gamma _{mn} a)}}N_m (\gamma _{mn} \rho )} \right]e^{ \pm im\phi } \cos \left( {p\pi \frac{z}{l}} \right)} } }\end{equation}

while the resonant frequencies are
    \begin{equation}\omega _{mnp}  = c\sqrt {\left( {\gamma _{mn} } \right)^2  + \left( {\frac{{p\pi }}{l}} \right)^2 } {\rm{  }}\end{equation}

with m=0,1,2,...  n=1,2,...  p=0,1,2,...

\section{Conclusions}

The presented formulae allow an exact calculus of the electric field
inside axial cavities with conducting walls, which may be used for a
rigorous evaluation of the radiation thermal flux in such a system.
For cryogenic systems, the numerical calculations need such accurate
formulae, taking into account that a large number of coaxial
insulators will be used and any errors will propagate and will be
recursively amplified. The only way to minimize the global errors is
to keep them as low as possible at any stage, thus involving from
the beginning these exact analytical solutions. That is why the
numerical calculus of the special functions involved must be
carefully conducted, implying an increased computing effort in order
to precisely describe the total radiation heat transfer rate in
cryogenic installations.

\appendix
\section*{APPENDIX}
\setcounter{section}{1}
\renewcommand{\theequation}{A.\arabic{equation}}

The Bessel functions $J_\nu  (x)$are solutions of Bessel equation
    \begin{equation}\left[ {\frac{{d^2 }}{{dx^2 }} + \frac{1}{x}\frac{d}{{dx}} + \left( {1 - \frac{{\nu ^2 }}{{x^2 }}} \right)} \right]R(x) = 0\end{equation}
and assume the following series expansions
    \begin{equation}J_\nu  (x) = \left( {\frac{x}{2}} \right)^\nu  \sum\limits_{j = 0}^\infty  {\frac{{( - 1)^j }}{{j!\Gamma (j + \nu  + 1)}}} \left( {\frac{x}{2}} \right)^{2j}\end{equation}

    \[J_{ - \nu } (x) = \left( {\frac{x}{2}} \right)^{ - \nu } \sum\limits_{j = 0}^\infty  {\frac{{( - 1)^j }}{{j!\Gamma (j - \nu  + 1)}}} \left( {\frac{x}{2}}
    \right)^{2j}\]

They satisfy the following orthogonality relation on the interval
$\rho  \in [0,a]$

    \begin{equation}\int\limits_0^a {\rho J_\nu  \left( {x_{\nu n} \frac{\rho }{a}} \right)} J_\nu  \left( {x_{\nu k} \frac{\rho }{a}} \right)d\rho  = \frac{{a^2 }}{2}\left[ {J_{\nu  + 1} \left( {x_{\nu n} } \right)} \right]^2 \delta _{nk}\end{equation}

The above solutions are independent for all $\nu $ except the case
of integer $\nu  = m = 0,1,2...$ when
    \[J_{ - m} (x) = ( - 1)^m J_m (x)\]

In this case a second independent solution of Bessel equation is the
Neumann function

    \begin{equation}N_\nu  (x) = \frac{{J_\nu  (x)\cos \nu \pi  - J_{ - \nu } (x)}}{{\sin \nu \pi }}\end{equation}

Both Bessel and Neumann functions as well as their combinations

\[H_\nu  ^{(1)} (x) = J_\nu  (x) + iN_\nu  (x){\rm{ }}\]

and

\[H_\nu  ^{(2)} (x) = J_\nu  (x) - iN_\nu  (x)\]

i.e the Hankel functions, satisfy the recursion formulae

\begin{equation}X _{\nu  - 1} (x) + X _{\nu  + 1} (x) = \frac{{2\nu }}{x}X _\nu  (x)\end{equation}

and

\begin{equation}X_{\nu  - 1} (x) - X_{\nu  + 1} (x) = 2\frac{{d X _\nu  (x)}}{{dx}}\end{equation}
where X(x) denotes a Bessel, Neumann or Hankel function.

\acknowledgements

This work was partially supported by the Romanian National Research
Authority (ANCS) under Grant 22-139/2008.

\section*{REFERENCES}

1)  J. D. Jackson, Classical Electrodynamics, 3-rd ed. John Wiley
and Sons, 1999 chapters 3 and 8.

2)  G.B. Arfken, H.J. Weber, Mathematical Methods for Physicists ,
6-th ed. Elsevier Inc. 2005, chapter 11.

3) L. D. Landau, E. M. Lifshitz and L. P. Pitaevskii,
Electrodynamics of Continuous Media", Pergamon Press
 (2nd edition, 1984).

\index{sco}

\end{document}